\documentstyle[12pt]{article}

\begin{document}
\begin{titlepage}
\title {\bf STRANGENESS PRODUCTION IN NEUTRON STARS }
\author{Sanjay K. Ghosh, S. C. Phatak and P. K. Sahu \\
Institute of Physics, Bhubaneswar-751005, INDIA.}
\maketitle
\begin{abstract}
Production of strange quarks in neutron stars is investigated in
this work. Three cases, one in which the energy and neutrinos produced
in the strangeness production reactions are retained in the reaction
region, second in which the neutrinos are allowed to escape the
reaction region but the energy is retained and the third in
which both the energy and neutrinos escape the reaction region
are considered. It is shown that the nonleptonic weak process
dominates strange quark production while semileptonic weak
processes, which produce neutrinos, lead to the cooling if the
neutrinos escape the reaction region. It is found that the time
required for the saturation of the strangeness fraction is between
$10^{-7}$ and $10^{-5}$ sec, with the shorter time corresponding
to the first two cases. About 0.2 neutrinos/baryon are emitted
during the process in the first two cases where as the neutrino
emission is somewhat suppressed in the last case. The average
energy of the neutrinos produced in all the three cases is found to
be several hundred $MeV$. We also find that a large amount of
energy is released during the strangeness production in the
first two cases and this leads to the heating of the reaction
region. Implications of the neutrino production are
investigated. \\
\end{abstract}
\end{titlepage}
\vfil \eject
\section{Introduction}

The lowest energy state of the quark matter should contain a
significant amount of strange quarks. For example, if we assume
equal mass u, d and s quarks, the strangeness fraction (defined
as the ratio of strange quark and baryon densities) in the quark
matter would be unity. Even if one considers a realistic value
of s quark mass ( about 150-200 MeV larger than u and d quark
masses\cite{bag} ), one can show that the strangeness fraction
in a chemically equilibrated quark matter is close to unity for
large enough baryon densities. Further more, it has been
conjectured\cite{Wit} that the quark matter with nonzero
strangeness fraction may even be the ground state of the dense
strongly interacting matter. This conjecture is supported by the
bag model calculations\cite{Farhi} for a certain range of s
quark mass and strong coupling constant.

On the other hand, the strangeness fraction in hadronic matter
(essentially the ratio of strange baryon and total baryon
densities) is expected to be small. Most of the hadronic
equations of state do not include strange baryons in the
hadronic matter\cite{eos} and the strangeness fraction for these
equations of state is zero. Nonzero strangeness fraction in the
hadronic matter can be generated by including strange baryons (
and their interactions ) in the calculations and/or by
considering the possibility of kaon condensation. Strange
baryons have been introduced by considering a Walecka model
Lagrangian density which includes hyperons and their
interactions with $\sigma$, $\omega$ and $\rho$
mesons\cite{Wal}. For a reasonable values of hyperon-meson
coupling constants, the strangeness fraction of hadronic matter
is found to be between 0.3 and 0.5\cite{GPS1} for baryon
densities between 0.3 and 0.6 $fm^{-3}$. The kaon condensation has
been considered by a number of authors\cite{Cley,Kap,Thor}. The
results of ref.\cite{Thor} show that, for baryon densities
between 0.4 and 1 $fm^{-3}$, the $K^-$ fraction (
which is same as the strangeness fraction ) varies from 0.4 to
0.9 as as one of the parameters ( $a_s m_s$ ) is varied from
-134 MeV to -310 MeV. This parameter depends on the strangeness
content of the nucleon ( varying from 0 to 0.2 ). Now, the
neutrino scattering experiments\cite{Ahren} indicate that the
strangeness content of the nucleon is $\sim 0.06$. It therefore
appears that, for reasonable value of the strangeness content of
the nucleon, the strangeness fraction due to the kaon condensate
could be about 0.5, a value similar to the one obtained in
Walecka model calculation\cite{GPS1}. In this regard, one must
note that hyperons are not included in the kaon condensation
calculations and for a reliable estimate of the strangeness
fraction in the presence of kaon condensate, this must be done.
Here we would like to point out that total kaon strangeness
fraction comes not only from the kaon itself but also from the
kaon-nucleon interaction term as suggested by Fuji et al.
\cite{Fuji94}. They find a monotonic increase of strangeness
number density with baryon density. But, here again the
strangeness fraction obtained remains uncertain due to
uncertainty inherent in the kaon-nucleon sigma term.

One expects that, if the baryon density in neutron stars is
large enough, the interior of the neutron star may consist of
the quark matter. If Witten's conjecture\cite{Wit} is correct,
the whole star will eventually become a strange star having
significant amount of the strangeness fraction. Otherwise, the
star will be a hybrid star with a core of strange quark matter.
Another interesting possibility, investigated by
Glendenning\cite{Glen}, is that, due to the presence of two
conserved charges ( the electronic and baryonic charges ) the
hybrid star may consist of a mixed phase region consisting of a
mixture of quark and hadronic matter. In any case, the quark
matter may be produced in the neutron star during the evolution
of the star, if the baryon density in the neutron star is large
enough. And when this occurs, significant amount of strange
quarks will be produced.  In the first case, the whole star will
eventually become a strange star. Even in case of the mixed
phase\cite{Glen}, macroscopic quark matter objects ( spheres,
rods or sheets\cite{objects} ) will be formed and strange quarks
will be produced in these objects.

Some of the mechanisms for the production of the strange quark
matter in neutron stars have been investigated by Alcock et
al.\cite{Alc}. These authors have assumed that Witten's
conjecture\cite{Wit} holds and proposed that the production of
the strange quark matter is initiated by an introduction of a
stable strange matter seed in the neutron star. This seed then
grows by `eating up' baryons in the hadronic matter, eventually
converting the neutron star to a star containing strange quark
matter. The speed at which the `conversion front', the boundary
between quark and hadronic matter, moves in a neutron star has
been calculated by Olinto\cite{Oli}, Olesen and Madsen\cite{OM}
and Heiselberg et al.\cite{Hei}. These calculations suggest that
the speed of the conversion front is between $ 10 \; m/sec$ and
$ 100 \; km/sec$. This means that the time required for the
conversion of the neutron star to the quark star is between $0.1
\; sec$ to $10^3 \; sec$. In this scenario, the strange quark
production takes place in the conversion front, which has a
microscopic thickness ( of few tens to few hundreds of fm ) in
comparison with the size of the star.

Collins and Perry\cite{CP}, on the other hand, have assumed that
the hadronic matter first gets converted into the
(predominantly) two-flavour quark matter, which later decays
into the three-flavour quark matter by weak processes.
Presumably, this is what happens when the mixed phase\cite{Glen}
is formed. That is, when the mixed phase is formed, the quark
matter objects are spontaneously generated and strangeness
production takes place in these objects. If the size of these
objects is large enough, one can use plane wave states to
calculate the strangeness production. Further more,
Lugones et al.\cite{Lug} have argued that, even when one
considers the production of the strange matter by
seeding\cite{Alc}, the conversion process may proceed as a
detonation instead of being a slow combustion process as assumed
in the preceding paragraph. This is partly due to the
instabilities of the conversion front under
perturbations\cite{Ben1}. In fact, it has been argued that such
a detonation may be responsible for the type II supernova
explosions\cite{Ben2}. In such situations, the strangeness
production would be occurring in an extended volume of the star.

Although, in both of these scenarios, the weak interactions are
responsible for the strangeness production, the detailed evolution
of the strangeness fraction is expected to be quite different.
In the former case, the production occurs in the conversion
front, a thin shell having a thickness of few tens of fm,
whereas in the later it occurs over an extended volume.
Thus, in the former case, the reaction products, the neutrinos,
and the energy generated during the strangeness production are
likely to escape the reaction region. On the other hand, in the
latter case, the heat generated may be trapped in the reaction
region if the thermal conductivity of the quark matter is not
large enough. This would lead to the heating of the reaction
region. Also, the neutrinos produced in the reaction region may
be trapped, as their mean free path is rather small in
comparison with the dimensions of the reaction region. This
means that the reaction rates are expected to be different for
the two scenarios.

Recently, Dai et al.\cite{Dai} have calculated the time required
for the strangeness production in quark matter. Unfortunately,
their results are not reliable as incorrect reaction rates have
been used in these calculations. Also, these authors do not
consider the trapping of energy and neutrinos in the reaction
region. In this paper, we want to present a detailed calculation
in which these aspects are included. In our calculation, the
reaction rates are determined from the basic weak interactions
and the time required for the chemical equilibration of the
quark matter has been calculated. We have also estimated the
number of neutrinos produced and the energy liberated during
chemical equilibration process. Our calculations show that the
saturation of the strangeness fraction is reached in $10^{-7} \;
- \; 10^{-5} \;$ sec. We also find that approximately 0.2
neutrinos/baryon are emitted during the equilibration process
and the average energy of these neutrinos is between 150 and 200
MeV. If we assume that the energy generated during the
strangeness production is trapped in the reaction region, the
temperature of the reaction region increases to about 50 MeV.

The paper is organised as follows. The reaction rates are
calculated in Section II. The results of the calculation are
presented and discussed in Section III. Section IV is devoted to
the conclusions.

\section{Reaction Rates}

The dominant reaction mechanism for the strange quark production
in quark matter is the nonleptonic weak interaction process;
\begin{eqnarray}
u_1 + d \leftrightarrow u_2 + s. \label{eq:nonlept}
\end{eqnarray}
Initially, when the quark matter is formed, the d-quark chemical
potential is much larger than s-quark chemical potential and the
reaction in eq(\ref{eq:nonlept}) converts d quarks into s
quarks, releasing energy. The amount of energy released depends
on the difference between $d$ and $s$ quark chemical potentials.
When the two chemical potentials are almost equal, the net
reaction rate due to the nonleptonic reaction becomes
vanishingly small. The nonleptonic reaction alone cannot produce
the chemical equilibrium since the chemical equilibration
requires that the d-quark chemical potential must be equal to
the sum of u-quark and electron chemical potentials. Thus
semileptonic weak interactions,
\begin{eqnarray}
d(s) \rightarrow u + e^- + \overline \nu_e~~
\end{eqnarray}
and
\begin{eqnarray}
u + e^- \rightarrow d  ( s ) + \nu_e,
\end{eqnarray}
must be included. In addition, we find that semileptonic
processes involving positrons,
\begin{eqnarray}
d(s)+ e^+ \rightarrow u + \overline \nu_e;~~
\end{eqnarray}
and
\begin{eqnarray}
u \rightarrow d  ( s ) +  e^+ +\nu_e,
\end{eqnarray}
must also be included. The reason being that, near the chemical
equilibrium, the electron chemical potential is rather small (
few MeV ) and the temperature of the reaction region is expected
to rise to few tens of MeV during the equilibration process. So
the electrons are not degenerate and therefore, the processes
involving the positrons must be included. It turns out that the
semileptonic reactions do not contribute much to the strangeness
production as such but these are responsible for the neutrino
production. In the following, the constituents of the reaction
region ( u, d and s quarks and electrons) are assumed to be in
thermal equilibrium at all times. This is reasonable since the
strong and electromagnetic interactions are responsible for the
thermal equilibration and their time scales are much smaller
than the weak interaction time scale.

Before we go on to the calculation of the reaction rates, let us
first consider the implications of the two scenarios of the
strangeness production. If the strangeness production occurs in
a thin boundary between quark and hadronic matter, the energy
and neutrinos produced in weak reactions will leave the reaction
region. Thus, the reaction will occur at constant temperature.
When the strangeness production takes place in an extended
volume, the situation is much more complicated. Depending on the
thermal conductivity of the quark matter, the size of the
reaction region and the strangeness saturation time scale, part
of the energy produced will escape from the reaction region. As
we shall see later, the strangeness-fraction saturation time is
less than $10^{-5}$ sec. We therefore expect that most of the
energy produced during the strangeness production will be
trapped in the reaction region and this would lead to heating of
the reaction region. As for the neutrinos, their mean free path
$\lambda_\nu$, in the quark matter of three times the nuclear
matter density is estimated to be $0.5/E_{\nu}^2 \;
km$,\footnote[1] {The mean free path is estimated using
formula $\lambda_{\nu}=\frac{1}{n_{B}\sigma}$, where $n_{B}$ the
baryon density and $\sigma$ is the scattering cross section
($q\nu\rightarrow q\nu$).} where $E_\nu$ is the neutrino energy
in $MeV$. Thus, except for the low energy neutrinos ($E_\nu <$
few tenth of MeV ), the neutrinos will be stopped in the
reaction region. This means that the weak reactions would occur
in a neutrino background. Therefore following three cases of
strangeness production are considered for a detailed
calculation.
\begin{enumerate}
\item The energy and neutrinos produced in the reaction are
retained in the reaction region. The momentum distribution of the
neutrinos is assumed to be that of noninteracting Fermi gas.
\item The energy produced in the reaction is retained in the
reaction region but the neutrinos are assumed to escape.
\item All the energy produced is assumed to be lost to the
surroundings and the temperature of the reaction region is
maintained at a fixed value.
\end{enumerate}
It should be noted that in the first case, the internal energy
of the system remains constant. Thus, in this case, the
temperature of the reaction region increases monotonically
during equilibration process. We have assumed a
noninteracting Fermi gas distribution function for the neutrinos
in case 1. Strictly speaking, this is not correct since the
neutrinos are not expected to reach thermal equilibrium quickly,
as they interact through weak interactions. Further more, the
low energy neutrinos can leave the reaction region, thus
altering the distribution function. However, as we shall find
later, the results for the cases 1 and 2 are almost identical.
In the second case, part of the internal energy is lost to the
escaping neutrinos. In this case, the nonleptonic weak reaction
leads to heating and semileptonic weak reactions lead to cooling
of the reaction region.

The reaction rates of the strangeness production reactions are
calculated using standard weak interaction Lagrangian
\cite{weak}. For the second and third case mentioned above,
these rates are,
\begin{eqnarray}
R_{d \rightarrow s} &=& C_{ds} \int p_dp_{u1}p_{u2}p_s
dp_ddp_{u1}dp_{u2}dp_s \; \;
\delta(\epsilon_d+\epsilon_{u1}-\epsilon_{u2}-\epsilon_s) \nonumber \\
& \times & \;\; {1 \over e^{(\epsilon_d-\mu_d)/T}+1} \; \; {1 \over
e^{(\epsilon_{u1}-\mu_{u1})/T}+1} \; \; {1 \over
e^{(\mu_{u2}-\epsilon_{u2})/T}+1} \; \; {1 \over
e^{(\mu_s-\epsilon_s)/T}+1} \nonumber \\
&\times& \int_{max \{ |p_d-p_{u1}| , |p_s-p_{u2}|
\} }^{min \{ |p_d+p_{u1}| , |p_s+p_{u2}| \}} \; dP \; \;
(1 + { p_d^2 + p_{u2}^2 - P^2 \over
2\epsilon_d \epsilon_{u2}}) \; \; (1 + { p_s^2 + p_{u1}^2 -P^2 \over
2\epsilon_s\epsilon_{u1}})
\label{eq:nl} \\
R_{u \rightarrow d \; (u \rightarrow s)}(e^-) &=& C_{ud} (C_{us})\int
p_up_{e}p_{\nu}p_{d(s)} dp_udp_{e}dp_{\nu}dp_{d(s)}
\delta(\epsilon_u+\epsilon_{e}-\epsilon_{\nu}-\epsilon_{d(s)}) \nonumber \\
& \times& \;\; {1 \over e^{(\epsilon_u-\mu_u)/T}+1} \; \; {1 \over
e^{(\epsilon_e-\mu_e)/T}+1} \; \; {1 \over
e^{(\mu_{d(s)}-\epsilon_{d(s)})/T}+1} \nonumber \\
& \times & \int_{max \{|p_u-p_{e}| , |p_{d(s)}-p_{\nu}| \} }^{min \{
|p_u+p_{e}| , |p_{d(s)}+p_{\nu}| \}}dP \; \;
(1 + { p_u^2 + p_{\nu}^2 - P^2 \over 2\epsilon_u \epsilon_{\nu}}) \; \; (1
+ { p_{d(s)}^2 + p_{e}^2 -P^2 \over 2\epsilon_{d(s)}\epsilon_{e}})
\label{eq:sl} \\
R_{d \rightarrow u ( s \rightarrow u)}(e^-) &=& C_{du} (C_{su})\int
p_{d(s)}p_{u}p_{e}p_{\nu} dp_{d(s)}dp_{u}dp_{e}dp_{\nu}
\delta(\epsilon_{d(s)}-\epsilon_{u}-\epsilon_{e}-\epsilon_{\nu}) \nonumber \\
& \times & \;\; {1 \over e^{(\epsilon_{d(s)}-\mu_{d(s)})/T}+1} \; \;
{1 \over e^{(\mu_{u}-\epsilon_{u})/T}+1} \; \; {1 \over
e^{(\mu_e-\epsilon_e)/T}+1} \nonumber \\
& \times & \int_{max \{
|p_{d(s)}-p_{\nu}| , |p_u-p_{e}| \} }^{min \{ |p_{d(s)}+p_{\nu}| ,
|p_u+p_{e}| \}} dP \; \;
(1 - { p_{d(s)}^2 + p_{\nu}^2 - P^2 \over 2\epsilon_{d(s)} \epsilon_{\nu}})
\; \; (1 - {P^2- p_u^2 + p_{e}^2 \over 2\epsilon_u\epsilon_{e}})
\label{eq:decay}
\end{eqnarray}
where
$C_{ds}=\frac{18G_{F}^{2}sin^2\theta_{C}cos^2\theta_{C}}{(2\pi)^5}
$, $~~C_{ud}
=C_{du}=\frac{6G_{F}^{2}cos^2\theta_{C}}{(2\pi)^5}$, $~~C_{su}=
C_{us}=\frac{6G_{F}^{2}sin^2\theta_{C}}{(2\pi)^5}$, $G_F$ is the
weak decay constant and $\theta_C$ is the Cabibbo angle and
$\epsilon_i$ and $\mu_i$ are energies and chemical potentials of
the $i$-th species ($i=u,d,s,e$). We have taken quark-gluon
interaction upto first order in strong coupling
constant\cite{GPS2}. The rate $R_{s \rightarrow d}$ is
calculated by interchanging d and s quarks and the rates
involving positrons are calculated by replacing $\mu_e$ by
$-\mu_e$ in the Fermi functions. In the presence of the neutrino
background, case 1, the rate equations are modified by
multiplying neutrino and antineutrino Fermi distribution
functions in eq(\ref{eq:sl}) and eq(\ref{eq:decay})
respectively.

Approximate analytic results for these rates have been obtained
for the massless particles and for temperatures small compared with
chemical potentials appearing in
eqs(\ref{eq:nl},\ref{eq:sl},\ref{eq:decay}) above. For the massive
s quarks and when the strong interaction between quarks are
included, such approximate rates are not available. In this
regard, we would like to note that Dai  et al.\cite{Dai} use
reaction rates obtained by generalizing Iwamoto formulae for the
non-equilibrium cases \cite{Iwa}. Thus, the reaction rates in
\cite{Dai} are proportional to $T^5$, where as, when $\mu_d \neq
\mu_s$, then for the small $T$ the reaction rates should depend
on the chemical potential difference and should not vanish in
$T\rightarrow 0$ limit. As we shall see later, the difference
between our reaction rates and those of Dai et al. lead to
different conclusions regarding neutrino emission. To compute
the reaction rates on has to perform the integrals in the
eqs(\ref{eq:nl},\ref{eq:sl},\ref{eq:decay}). In these equations
the $P$-integral can be done analytically. The energy delta
function is used to perform one of the remaining four integrals.
As a result the eqs(\ref{eq:nl},\ref{eq:sl},\ref{eq:decay})
reduce to three dimensional integrals. These are evaluated
numerically using extended Simpson's rule. We have ensured that
the numerically calculated results agree with the approximate
analytic results for massless particles at small temteratures.
Note that the energy carried away by the neutrinos and
antineutrinos ( the neutrino emissivities $\epsilon_{\nu}$ and
$\epsilon_{\overline \nu}$ ) is calculated by including the
neutrino energy in the integrands of eq(\ref{eq:sl}) and
eq(\ref{eq:decay}) respectively.

The rate of change of densities of the constituents ( u, d, s
quarks and electrons) are obtained from the reaction rates given
above. These are,
\begin{eqnarray}
\frac {dn_u(t)}{dt} &=& R_{d \rightarrow u}(e^-) +
R_{s \rightarrow u}(e^-) - R_{u \rightarrow d}(e^-) - R_{u
\rightarrow s}(e^-) \nonumber \\
&+& R_{d \rightarrow u}(e^+) + R_{s \rightarrow u}(e^+) - R_{u
\rightarrow d}(e^+) - R_{u \rightarrow s}(e^+)
\label{eq:uden} \\
\frac {dn_d(t)}{dt} &=& R_{s \rightarrow d} - R_{d\rightarrow s} -
R_{d \rightarrow u}(e^-) + R_{u \rightarrow d}(e^-)\nonumber \\
&-& R_{d \rightarrow u}(e^+) + R_{u \rightarrow d}(e^+) \label{eq:dden}
\end{eqnarray}
The reactions in eqs(\ref{eq:nl},\ref{eq:sl},\ref{eq:decay})
satisfy charge and baryon number conservation. Thus the rate of
change of s quark and electron densities are known from
eq(\ref{eq:uden},\ref{eq:dden}). In addition, when the neutrinos
are allowed to escape the reaction region, the internal energy of the
system decreases and the rate of decrease of the internal energy
is given by,
\begin{eqnarray}
\frac {d\epsilon} {dt} = - \epsilon_{\nu} - \epsilon_{\overline
\nu} \label{eq:e}
\end{eqnarray}

The calculation of the evolution of the constituent densities
proceeds as follows. Intially, ( time t = 0 ), we assume that
the quark matter having same chemical composition as that of the
hadronic matter is formed. This gives the initial condition on
quark and electron densities and their chemical potentials. The
initial temperature is varied between 5 and 20 MeV. The internal
energy and number densities of the constituents at later time
are calculated by integrating the rate
equations(eqs(\ref{eq:uden},\ref{eq:dden},\ref{eq:e}))
numerically using Runge-Kutta method. Assuming the thermal
equilibrium, the chemical potentials of the constituents and the
temperature are determined at each instant of time. Note that
these are required for the calculation of the reaction rates and
neutrino emissivities. The calculation is continued till the
strangeness fraction saturates. The results of the calculation
are discussed in the following section.

\section{Results And Discussions}

The calculations have been performed for a range of baryon
densities ( 0.4 - 0.8 fm$^{-3}$), strange quark masses ( 150 - 200
MeV ) and strong coupling constants ( 0.1 - 0.5 ). For a
detailed discussion, the strange quark mass of 150 MeV and
the strong coupling constant of 0.1 has been chosen. The results for
other parameter sets are similar. The parameter set used by Dai
et al.\cite{Dai} ($n_B=0.4fm^{-3}$, strange quark mass of 200
MeV and strong coupling constant 0.17) has also been
included in the discussion. This baryon density corresponds to
the baryon density at the center of the neutron star of 1.4
M$_\odot$.

The knowledge of the initail and final values of energy/baryon (
or equivalently, the energy density ) could be of interest. For
example, for the parameter set ( $n_B = 0.4 fm^{-3}$, strange
quark mass = 150 MeV, strong coupling constant of 0.1 and
$B^{1/4}$ = 150 MeV ) the initial and final energy/baryon is 1100
MeV and 970 MeV respectively\footnote[2]{ The bag pressure term
does not affect the {\it difference} in the energy densities. In
any case, the bag energy is a small fraction of the total energy
density at these baryon densities}. Thus, 130 MeV of
energy/baryon is liberated during the strangeness production.
Part of this energy is carried by the neutrinos and rest is used
in heating of the reaction region. The latent heat produced
during the hadron-quark phase transition is, however, not
included in our calculation.

Let us now consider the first two cases. The time dependence of
the temperature and strangeness fraction are shown in Fig 1.
The figure shows that for both the cases the strangeness
fraction attains its saturation value in a very short time (
$\sim 10^{-7} \;$ sec ). The time required for the chemical
equilibration as well as the strangeness saturetion time is
about the same for the two cases. Although not shown here, the
reaction rates for the two cases are also almost same. This
implies that the strangeness production does not depend very
much on whether the neutrinos are stopped in the reaction region
or not. As for the temperature, the figure shows that, the
temperature of the reaction region rises to about $50 \; MeV$ in
$\sim 10^{-7} \; sec$ for case 1. After chemical equilibration,
the temperature of the reaction region remains constant. For the
second case, the temperature reaches a peak value of about $45
\; MeV$ after about $10^{-8} \; sec$ and later it continues to
fall. The peak value of the temperature is about 10\% smaller
than the temperature obtained in case 1. Initially, the time
dependence of the temperature is similar for the two cases. The
smaller rise of the temperature and the decrease of the
temperature at later times in the second case is because of the
loss of the internal energy to the neutrinos which are assumed
to escape the reaction region.

The reaction rates for the case 2 are plotted in Fig 2. We do
not show the reaction rates for the case 1 as these are
practically same as those of case 2. This figure shows that the
reaction rate of the nonleptonic weak process is several orders
of magnitude larger than the semileptonic rates. Thus, the
nonleptonic weak process dominates the strangeness production
till the strangeness fraction reaches saturation value.
Afterwards the nonleptonic weak rate drops sharply. It is
interesting to note that initially, the rates $R_{u \rightarrow
d}(e^-)$ and $R_{u \rightarrow s}(e^-)$ are much larger than
other semileptonic rates. This is because of the large initial
value of the electron chemical potential. At later times, when
the temperature reaches the peak value, $R_{d \rightarrow
u}(e^+)$ and $R_{s \rightarrow u}(e^+)$ are comparable to $R_{u
\rightarrow d}(e^-)$ and $R_{u \rightarrow s}(e^-)$
respectively. The behaviour of the reaction rates clearly shows
that the semileptonic processes do not contribute much to the
strangeness production. Initially, the primary mechanism of the
strangeness production is the nonleptonic process in which d
quarks are converted to $s$ quarks. Later, when the semileptonic
rates are comparable to the nonleptonic rate, the semileptonic
reactions do not alter the $s$ quark density very much since the
strangeness fraction has already saturated. In other words, the
primary mechanism of $s$ quark production is the nonleptonic weak
process with semileptonic weak processes playing a minor role.
The semileptonic processes, however, contribute to the neutrino
production. This also leads to the cooling of the reaction
region in case 2.

The reaction rates and strangeness fraction when temperature is
kept at 10 $MeV$ are plotted in Fig 3. In this case also the
nonleptonic weak interaction dominates. The figure shows that
the time required for the saturation of strangeness fraction is
larger ( $ \sim 10^{-5} \; sec $ ). Furthermore, the time
required for the chemical equilibration is even larger ( about 0.1
sec ). This is because of the small values of the rates $R_{u
\rightarrow d}$ and $R_{u \rightarrow s}$. Thus we find that the
chemical potentials of $d$ and $s$ quarks become equal earlier
but the chemical equilibrium ( $\mu_d (\mu_s) = \mu_u + \mu_e$ )
is reached at much later time. Similar results are obtained for
the temperature of the reaction region ranging between 5 and 20
$MeV$. It must be noted that the choice of the temperature in
this case is somewhat uncertain. We have noted earlier that,
when the energy is retained in the reaction region, the initial
rise in the temperature of the reaction region is rapid. For
example, figure 1 shows that the temperature rises to 20 $MeV$
in $10^{-9} \; sec$.  Therefore, it is reasonable to assume a
temperature of 10 - 20 $MeV$ in the reaction region. The
calculations show that the strangeness saturation time decreases
with the increase in the temperature.

The results for different baryon densities, and for different
values of the initial strangeness fraction are summarised in
Table I. Case I and II correspond to zero and nonzero initial
strangeness fraction respectively. The initial value of
strangeness fraction in case II is obtained from Walecka model
calculations \cite{GPS1}. Case III corresponds to the parameter
set of \cite{Dai} and the case IV corresponds to the situation
when temperature is held fixed. Table I shows that the
saturation time increases by orders of magnitude when the
temperature of the reaction region is held constant. The time
required for the saturation decreases when the initial
strangeness fraction is nonzero and when the baryon density is
increased. From the results displayed in Table I, one can
conclude that the time required for the saturation of
strangeness fraction varies between $10^{-7}$ and $10^{-5}$ sec,
with the larger time corresponding to the constant temperature
case.

The number of neutrinos emitted during the strangeness
production and their average energy is also shown in Table I. We
find that 0.07 to 0.25 neutrinos/baryon are emitted during
strangeness production and average energy per neutrino is
between 150 and 200 $MeV$.  The neutrino emission is smaller for
the constant temperature case and at lower temperatures, the
neutrino production further decreases. However, in constant
temperature case, the neutrinos emitted after the saturation of
the strangeness fraction are not included in the table. Thus,
the actual neutrino number, in this case, would be somewhat
larger than the values shown in Table I. The average energy of
neutrinos appears to be rather large. This is because the
neutrino energy is essentially related to the difference between
$ |\mu_u + \mu_e - \mu_d ( \mu_s )|$, which is few hundred $MeV$
in the begining. In contrast to our results, Dai et
al.\cite{Dai} predict much more copious neutrino production with
each neutrino carrying about 0.1 $MeV$ of energy on the average.
In their calculations, they use Iwamoto like formulae\cite{Iwa}
in which the average neutrino energy is proportional to the
temperature. Since they choose a temperature of 0.1 $MeV$, the
average energy of the neutrinos in their calculations is $\sim $
0.1 $MeV$.

Let us now consider the consequences of strangeness production
in a neutron star. Consider the scenario in which the
strangeness production takes place in a thin conversion front.
Assuming that the whole star of mass 1.4 M$_\odot$ is converted
into a strange star ( a typical mass of a neutron star ), about
$10^{56}$ neutrinos will be produced during the conversion and
the energy carried by these neutrinos will be about $2.5 \times
10^{52}$ ergs. In addition, a significant amount of energy will
be produced in the form of heat during the conversion process. A
rough estimate suggests that this would be about the same order of
magnitude. Now, the time during which these neutrinos are
produced is not given by the strangeness production time
displayed in Table I, but is related to the time taken by the
conversion front to traverse the star. This time has been
estimated to be between $0.1$ and $10^3$ sec \cite{Oli,OM,Hei}.
The neutrinos produced during the conversion process cannot
leave the star as their mean free path is much smaller than the
radius of the star. For example, the mean free path of neutrinos
in neutron and quark matter of three time nuclear matter density
is about $0.1/E_{\nu}^2$ km\cite{Be} and $0.5/E_{\nu}^2$ km
respectively. Thus, these neutrinos will loose their energy
while percolating out of the star and during the process, more
neutrino-antineutrino pairs will be produced. The situation is
similar to what happens during the supernova
explosion\cite{Bah}. The supernova calculations suggest that
roughly 10 pairs of neutrinos are produced during percolation.
Since the energy of neutrinos produced during the strangeness
production is much larger ( $\sim 200 \; MeV$ {\it vs} $ \sim 15
\; MeV$ in supernova collapse ), many more neutrino pairs will
be produced. Thus, the number of neutrinos eventually escaping
the star may be $10^{57}$ or larger.

Now consider the scenario in which the strangeness production
takes place in an extended volume. In this case, only a part of the
star is likely to be converted into the strange matter. As an example,
let us assume that the mass of the strange quark matter formed
in such a process is about 0.7 M$_\odot$. Then about
$2 \times 10^{56}$ neutrinos will be emitted in $10^{-7} \;
sec$. The energy carried by these neutrinos will be about
$5 \times 10^{52} \; ergs$. Also, the quark matter will be
heated to about 50 $MeV$ of temperature. In this case also, the
neutrinos produced will percolate outwards, producing
neutrino-antineutrino pairs in the process. In contrast, the
standard supernova calculations, in which quark matter formation
is not included, predict that $\sim 10^{57}$ neutrinos produced
and the average energy of these neutrinos is about $15 \;
MeV$\cite{Bah}. So, if quark matter is formed during the
supernova collapse, the number of neutrino produced and the
energy carried by these neutrinos will possibly be larger by an
order of magnitude. Now, it is known that the prompt shock model
\cite{BBB}, which is based on the hydrodynamical bounce, does
not produce sufficient bounce for the supernova explosion\cite{BL}.
Hence, a delayed mechanism, in which heating by an intense
neutrino flux coming from the collapsed core\cite{BW}, has been
proposed as one of the mechanisms of supernova explosions. If
the strange matter is formed the during supernova collapse, the
neutrinos produced during strangeness production would provide a
significant addition to the neutrino flux, thus enhancing the
delayed mechanism. Such a scheme has been proposed by Benvenuto
and Hovarth\cite{Ben2}.

{}From the preceding discussion, one can conclude that the
strangeness production in a dense star will be followed by
copious production of neutrinos and release of large amount of
energy. It may be possible to observe the neutrinos produced in
the strangeness production process. Particularly, if the
strangeness production occurs in a collapsing supernova, the
neutrinos produced may help in generating the shock required for
the supernova explosion. However, it must noted that a detailed
calculation of neutrino transport in quark and nuclear matter is
required before quantitative estimates can be made.

\section{Conclusions}

The strangeness production when a neutron star gets
converted to a strange star is studied in this work. In our
calculation we consider three cases. These are a) the neutrinos
and energy produced during the strangeness production are
retained in the reaction region, b) the energy is retained in
the reaction region but the neutrinos are allowed to escape and
c) both the energy and neutrinos are allowed to escape. The
first two correspond to the scenario in which the strangeness
production occurs in an extended volume where as the last case
corresponds to the scenario in which the production takes place
in a thin conversion front. We find that, the strangeness
production occurs in $10^{-7}$ sec in the first two cases and
$10^{-5}$ sec in the last case. A large amount of energy is
produced during the production and this may raise the
temperature of the reaction region to about 50 $MeV$. Thus, quark
matter formation in a neutron star is a violent process in which
a large amount of energy is released. The main reason for this
is the large difference in the strangeness fraction of nuclear
and quark matter ( 0.0 - 0.5 for the nuclear matter {\it vs} $\sim$
1.0 for the quark matter ). About 0.2 neutrinos/baryon are
emitted during this process and the average energy of these
neutrinos is hundreds of $MeV$. Although these neutrinos cannot
escape the star as their mean free path in hadronic and quark
matter is rather small, these will percolate out, loosing their
energy and producing large number of neutrino-antineutrino pairs
in the process. It may be possible to detect these neutrinos.
But a detailed calculation of neutrino transport in quark and
nuclear matter is required to estimate the flux of these
neutrinos.
\vfil
\eject

\vfil
\eject
\newpage
\begin{table}
\caption {The values of time required for the strangeness fraction
saturation, energy/neutrino (Av. Energy), total neutrinos per
baryon ($N_{{\nu}B}$) and total neutrinos ($N_{\nu}$) emitted from a star
of mass $1.4 M_\odot$ for four different cases are displayed.}
\hskip 0.5 in
\begin{tabular}{ccccccc}
\hline
\multicolumn{1}{c}{$n_B$} &
\multicolumn{1}{c}{$T$} &
\multicolumn{1}{c}{$Time$} &
\multicolumn{1}{c}{$Av. Energy$} &
\multicolumn{1}{c}{$N_{{\nu}B}$} &
\multicolumn{1}{c}{$N_{\nu}$} &
\multicolumn{1}{c}{$Cases$}\\
\multicolumn{1}{c}{($fm^{-3}$)} &
\multicolumn{1}{c}{($MeV$)} &
\multicolumn{1}{c}{($seconds$)} &
\multicolumn{1}{c}{($MeV$)}&
\multicolumn{1}{c}{}&
\multicolumn{1}{c}{} \\
\hline
0.45&--&2.41$\times 10^{-7}$&168.5&0.24&3.75$\times 10^{56}$& \\
0.60&--&1.21$\times 10^{-7}$&194.0&0.27&4.22$\times 10^{56}$& (I)\\
\hline
0.45&--&2.47$\times 10^{-7}$&162.6&0.17&2.66$\times 10^{56}$& \\
0.60&--&1.12$\times 10^{-7}$&182.9&0.15&2.34$\times 10^{56}$& (II)\\
\hline
0.40&--&2.22$\times 10^{-7}$&176.1&0.25&3.91$\times 10^{56}$& (III)\\
\hline
0.60&20.0&3.14$\times 10^{-6}$&140.0&0.07&1.09$\times 10^{56}$& \\
0.60&10.0&6.70$\times 10^{-5}$&185.4&0.03&0.47$\times 10^{56}$& (IV)\\
0.60&5.0&6.82$\times 10^{-5}$&227.9&0.008&0.12$\times 10^{56}$& \\
\hline
\end{tabular}
\end{table}
\vfill
\eject
\newpage
\begin{figure}
\caption { Time dependence of temperature and strangeness
fraction are shown for density $0.6~fm^{-3}$, (a) and (c)
correspond to temperature and strangeness fraction for case 2
and (b) and (d) for case 1 respectively.}
\vskip 0.4in
\caption { The reaction rates of nonleptonic and semileptonic
weak processes till the strangeness fraction reaches saturation value
are shown in lower half, where (a) is $u +e^- \rightarrow d + \nu_e$, (b)
is $u + e^- \rightarrow  s  + \nu_e$, (c) is $d\rightarrow s$, (d) is
$d+ e^+ \rightarrow u + \overline \nu_e$ and (e) is $s
+e^+\rightarrow u + \overline \nu_e$. Here $n_B=0.6~fm^{-3}$ with
nonzero initial strangeness. The other semileptonic reaction
rates are much lower and are not shown here. }
\vskip 0.4in
\caption { The upper half of the figure is strangeness fraction for
fixed temperature $10~MeV$. The
reaction rates of nonleptonic and semileptonic weak processes till the
strangeness fraction reaches saturation value are shown in lower half,
where (a) is $u +e^- \rightarrow d + \nu_e$, (b)
is $u + e^- \rightarrow  s  + \nu_e$, (c) is $d\rightarrow s$ and (d) is
$u \rightarrow s + e^+ +\nu_e$. Here $n_B=0.6~fm^{-3}$ with
nonzero initial strangeness. The other semileptonic reaction
rates are much lower and are not shown here.}
\end{figure}
\vfil
\eject
\end{document}